\documentclass[twocolumn,preprintnumbers,amsmath,amssymb,showpacs,floatfix]{revtex4}

\usepackage{graphicx}
\usepackage{dcolumn}
\usepackage{bm}

\begin{document}

\title{Ultra-low-phase-noise cryocooled microwave dielectric-sapphire-resonator oscillators with $1\times10^{-16}$ frequency instability}

\author{John G.~Hartnett,$^1$, Nitin R.~Nand,$^1$ and Chuan~Lu$^2$}

\email{john.hartnett@uwa.edu.au}
\affiliation{$^1$School of Physics, University of Western Australia, 35 Stirling Hwy, Crawley 6009 WA, Australia\\
$^2$School of Information Science and Technology, West Campus, University of Science and Technology of China, Hefei, Anhui, 230027, P.R. China}

\date{\today}

\begin{abstract}

Two nominally identical ultra-stable cryogenic microwave oscillators are compared. Each incorporates a  dielectric-sapphire resonator cooled to near  6 K in an ultra-low vibration cryostat using a low-vibration pulse-tube cryocooler.  The phase noise for a single oscillator is measured at -105 dBc/Hz at 1 Hz offset on the 11.2 GHz carrier. The oscillator fractional frequency stability is characterized in terms of Allan deviation by $5.3\times10^{-16}\tau^{-1/2}+9 \times 10^{-17}$ for integration times $0.1 \, s < \tau < 1000\, s$ and is limited by a flicker frequency noise floor below $1\times10^{-16}$. This result is better than any other microwave source even those generated from an optical comb phase-locked to a room temperature ultra-stable optical cavity.  
\end{abstract}

\pacs{77.90.+k, 65.40.-b, 06.30.Ft, 07.20.Mc}


\maketitle

The time and frequency standards community is ever advancing the state-of-the-art in both stability and accuracy of atomic clocks \cite{hartnett2011}. With the development of atomic fountain clocks it was found that a more stable microwave interrogation oscillator was required than was available from an ultra-stable quartz oscillator to avoid the Dick effect \cite{Dick1987, Dick1990,Audion1998, Santarelli1998}. The latter results from the fact that most atomic fountain clocks have significant dead time in their interrogation cycle, simply due to the fact that they are pulsed devices
where one has to wait until the detection process is finished before the next cloud of cold atoms is launched \cite{Wynands2005}. This results in the introduction of the phase noise of the local reference oscillator, from which the microwave cavity frequency is derived, to the short-term stability of the fountain clock. The problem was overcome through the use of cryogenic sapphire oscillators with significantly much better short-term frequency stability \cite{Chang2000} and consequently it has resulted in atomic fountain clocks reaching a performance limited only by quantum projection noise \cite{Santarelli1999}.  

These microwave oscillators based on a dielectric-sapphire resonator cryogenically cooled to nearly 6 K are in use in several standards labs and have been found to be useful for various applications \cite{Wolf, Stanwix, Marra2007, Watabe2007}. The oscillators \cite{Hartnett2006, Locke2008, Tobar2006}, until recently, have relied on large dewars  of liquid helium to cool the resonator with a few exceptions \cite{Dick2000, Wang2004, Watabe2003, Watabe2004}  where a cryocooler was used. However the latter had their short-term performance compromised by vibrations from the cryocooler itself. More recently low vibration pulse-tube cryocoolers and specially designed cryostats \cite{Wang2010} have enabled cryogenic oscillators as equally stable \cite{Hartnett2010a, Hartnett2010b, Grop2010a, Grop2010b, Nand2011} yet without being limited by the noise from the  cryocooler. 

The present design  is currently state-of-the-art. A silver-plated copper cavity encloses a 51 mm diameter and 30 mm high HEMEX grade sapphire cylindrical resonator and is the same as previously reported \cite{Hartnett2006} (see Fig. 1 therein) but the thermal design for housing it in the cryostat is significantly different \cite{Hartnett2010b}. See Fig. 1. Also long coaxial cables connecting the cryogenic resonator to the room temperature loop oscillator (feedback amplifier and microwave electronics) are no longer used as the cryostat is very compact and as a result the oscillator does not suffer from a slowly decreasing level of liquid helium as it does in the case with large liquid helium dewars.  The cryostat \cite{Wang2010}  uses only about one liter of liquid helium that is constantly reliquified in a closed system by the low-vibration cold head. There is no hard contact between the condenser and the dielectric resonator resulting in significant reduction in vibrations and an enhanced temperature stability at the resonator due to specially engineered thermal filtering. The liquid helium cools the cold finger to which the  vacuum can containing the sapphire resonator is attached.

\begin{figure}[!t]
\centering
\includegraphics[width=3in]{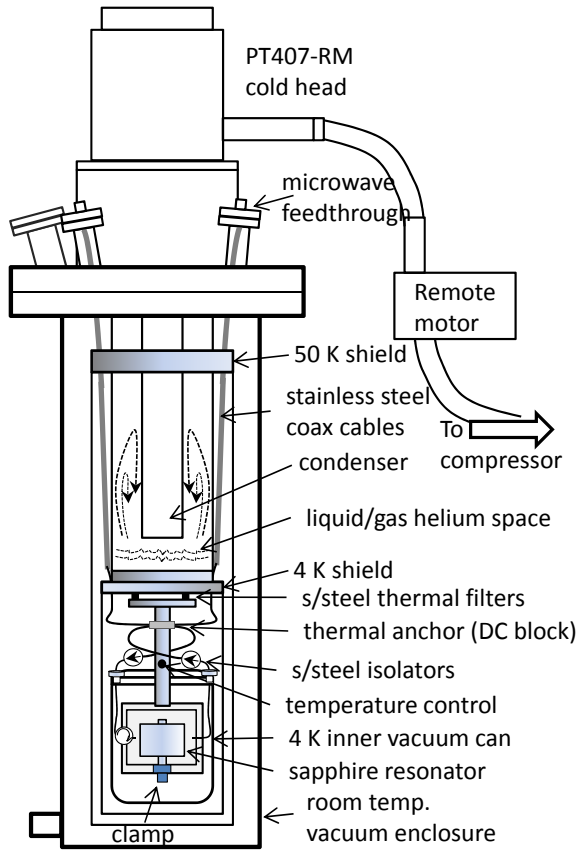}
\caption{Schematic of the CryoMech low-vibration cryostat \cite{Wang2010} housing the sapphire resonator on the cold finger cooled by about a liter of liquid helium in the enclosed liquid/gas helium space. The CryoMech PT407-RM pulse-tube condenser constantly maintains the liquid helium bath near 4 K. A heater, sensor and a Lakeshore 340 temperature controller maintains the temperature at the sapphire crystal within about 10 $\mu$K in the short-term by controlling the temperature at the resonator frequency-temperature turnover point.}
\label{Fig1}
\end{figure}

For the first time two cryogenic oscillators of this design were implemented with output frequencies near 11.2 GHz but differing by 1.58 MHz. At the control temperature for each resonator near 6 K, the resonators have loaded Q-factors of 1 billion and primary coupling better than 0.8 via loop probes made from the coaxial lines  connecting to the room temperature loop oscillator. This paper reports on the inferred phase noise and frequency stability of a single oscillator determined from a comparison of two nominally identical and independent oscillators using this type of cryostat and a pulse-tube cryocooler. Previously comparisons were made against a second oscillator using a large liquid helium dewar \cite{Hartnett2010a, Hartnett2010b}. 

In order to compare the oscillators which output at slightly different microwave frequencies near 11.2 GHz, a technique was implemented that could measure both their phase noise and frequency stability in real-time. With this technique the noise performance of the oscillators could be easily optimized and at  power levels incident on the resonator, much lower than had previously been used, better frequency stability was achieved.

\textit{Experimental Method}-- A block diagram of the measurement technique which enabled simultaneous measurement of the relative phase noise and frequency stability of two  cryocooled oscillators is shown in Fig. 2. The 11.201967 GHz signal from one of the cryogenic oscillators, cryoCSO1, was split via a 3-dB power splitter and one of the outputs was down-converted to approximately 100 MHz via a 112 divider chain. The chain was constructed from the series connection of three low noise dividers (successive division by 2, 4 then 14).  The 100 MHz output was then used as the very low phase noise reference signal in a Symmetricom 5125A test set. The input signal to the test set is the 1.58 MHz beat note of the two oscillators derived from the output of the microwave mixer after low pass filtering. 

The test set outputs the single sideband (SSB) phase noise ${\mathcal L}_{\varphi}(f)$ in units dBc/Hz and the Allan deviation of the fractional frequency fluctuations of the input signal relative to the reference signal \cite{testset}. The particular model used has an input  frequency range from 1 MHz to 400 MHz, therefore it was necessary to down-convert from the microwave frequency to enable the comparison. In addition the beat note was also sampled with an Agilent 53132A $\Lambda$-type frequency counter \cite{Dawkins} with a gate time of 10 s, and a computer controlled data acquisition system. 

\begin{figure}[!t]
\centering
\includegraphics[width=3in]{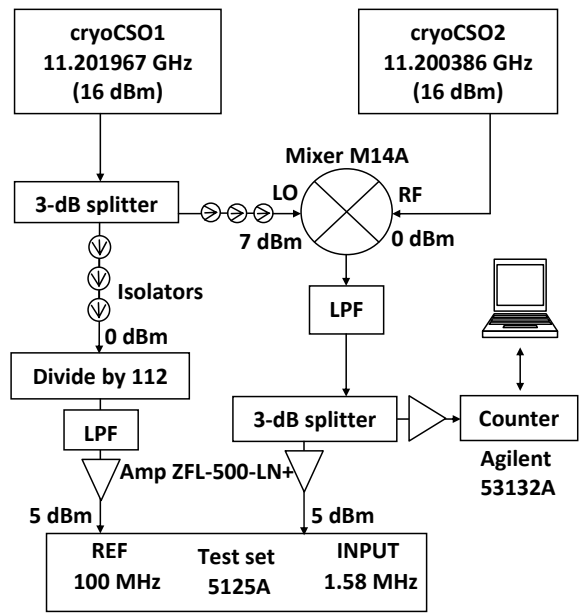}
\caption{A block diagram of the measurement technique used to compare two cryocooled oscillators (cryoCSO1 and cryoCSO2) with a Symmetricom 5125A phase noise test set (attenuators and DC blocks not shown). The principle components used were; a Watkins Johnson M14A mixer, Mini-Circuits amplifiers ZFL-500LN+, and a divide by 112 chain. The latter was constructed from the series connection of Hittite dividers: the HMC364 divide by 2, the HMC365 divide by 4, and the HMC705LP4 programmable divider which was set to divide by 14. LPF = Low Pass Filter.}
\label{Fig2}
\end{figure}

From a prior comparison of one cryocooled sapphire oscillator with a liquid helium cooled version it was expected that the phase noise for a single oscillator would be about -97 dBc/Hz at 1 Hz offset \cite {Hartnett2010a, Hartnett2010b}. Dividing down the 11.2 GHz frequency by the factor 112 means a 41 dB reduction in phase noise and hence the phase noise of the 100 MHz reference from the output of  the divider chain is dominated by the residual phase noise of the dividers themselves. This was confirmed by measuring two nominally identical divider chains with independent cryocooled oscillators at their inputs. Their total phase noise was the same as their residual phase noise (see curve 1 of Fig. 3), which is about 30 dB lower than the phase noise of the 11.2 GHz signal of the microwave oscillator. This means curve 1 of Fig. 3 is the noise floor of our phase noise measurement system.

\begin{figure}[!t]
\centering
\includegraphics[width=3in]{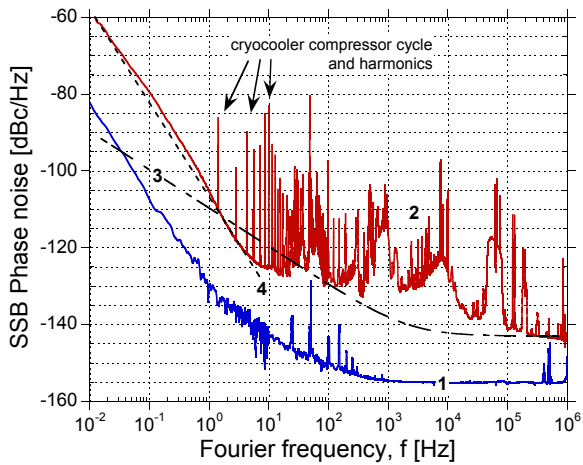}
\caption{(color online) The SSB phase noise of the 100 MHz divider chain (curve 1), and the SSB phase noise for a single oscillator at 11.2 GHz (curve 2). Curve 3 represents the phase noise  of the Endwave JCA microwave amplifier used in the loop oscillator,  measured at 11.2 GHz with an incident power level of -30 dBm, which is approximately the loop oscillator operating condition. Curve 4 is a best fit for the phase noise of the locked oscillator for Fourier frequencies $f\leq4$ Hz.}
\label{Fig3}
\end{figure}

\begin{figure}[!t]
\centering
\includegraphics[width=3in]{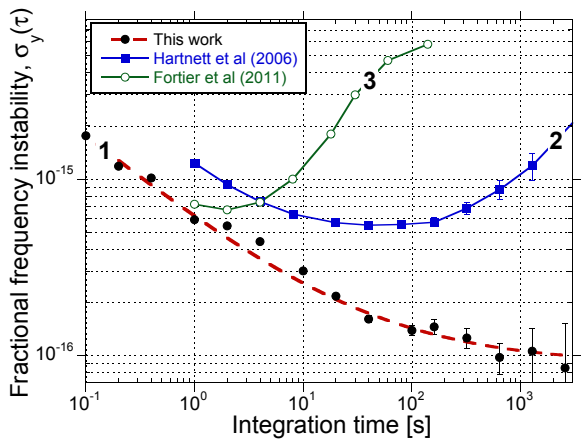}
\caption{(color online) The fractional frequency stability (Allan deviation $\sigma_y(\tau)$) for a single 11.2 GHz crycooled  sapphire oscillator as a function of integration time, $\tau$, (curve 1). The broken line fit is described by Eq. (\ref{stab}). Curve 2 is the stability for a single 11.2 GHz cryogenic oscillator using the same resonators as in this paper but cooled in a large liquid helium bath \cite{Hartnett2006}. Curve 3 is the fractional frequency stability of a 10 GHz microwave signal derived the NIST photonic generator \cite{Fortier}. The values for $\sigma_y(\tau)$ where $\tau>10$ s of curve 1 and all of curve 2 are estimated from a $\Lambda$-counter after a linear frequency drift was removed. Curve 3 includes drift, but its impact on stability for $\tau < 100$ s is negligible. }
\label{Fig4}
\end{figure}

\textit{Results and discussion}-- The measured SSB phase noise of a single cryocooled sapphire oscillator is shown as curve 2 of Fig. 3 where the power incident on the resonator is 0.1 mW.  Also the measured phase noise of the high gain  microwave amplifier, used in the loop oscillator, is shown by the broken line (curve 3). The latter was measured out of the loop with an input power of -30 dBm, which is approximately its in-loop operating condition. 
The phase noise for a single oscillator, at Fourier frequencies $f\leq$ 4 Hz, can be approximately described by the fit (curve 4) in Fig. 3, and is given by
\begin{equation}
{\mathcal L}_{\varphi}(f) = 10\;log\left(10^{-11.7}f^{-3}+10^{-10.9}f^{-2} +10^{-14.6}\right),
\end{equation}
[dBc/Hz]. This is characteristic of white frequency ($f^{-2}$) noise at Fourier frequencies $f \leq 4$ Hz and flicker frequency ($f^{-3}$) noise at much lower frequencies. The measured SSB phase noise of -105 dBc/Hz at 1 Hz offset is state-of-the-art for any microwave oscillator, only 5 dB above that of the high gain microwave amplifier. For Fourier frequencies 1 Hz $\leq f \leq$ 10 Hz this result is comparable with the SSB phase noise of a 10 GHz signal generated from an optical comb phase-locked to an ultra-stable room temperature optical cavity -- the NIST photonic generator \cite{Fortier}. 

Outside the bandwidth of the Pound frequency control servo (of several hundred Hz) the cryogenic sapphire oscillator is limited by the phase noise of the microwave amplifier which reaches a thermal phase noise floor near -146 dBc/Hz at Fourier frequencies $f > 10^{5}$ Hz. The spurs seen in Fig. 3 are largely due to the 1.41 Hz cryocooler compressor cycle frequency, the 50 Hz mains power, and their harmonics and near 70 kHz  Pound modulation sidebands. Ignoring the spurs the best fit to the phase noise for $f>4$ Hz is described by ${\mathcal L}_{\varphi}(f) = 10\;log(10^{-12.1}f^{-0.45}+10^{-14.6})$ [dBc/Hz].

The oscillator frequency stability was evaluated for various levels of microwave power ($P_{inc}$) incident on the cryogenic resonator and at $P_{inc}$ = 0.1 mW the best stability was achieved. Fig. 4 shows the frequency stability for a single oscillator, in terms of Allan deviation ($\sigma_y(\tau)$),   after a linear fractional frequency drift of about $4.4\times10^{-14}$/day was subtracted, and data collected for at least 3 hours. At the time the oscillators had been running for only between 1 to 2 weeks. (It is known that their longer term stability improves with aging \cite{Nand2011}.)  The oscillator stability (curve 1)  is described by,
\begin{equation}
\label{stab}
{\sigma}_{y}(\tau) = 5.3\times10^{-16} \tau^{-1/2} + 9\times10^{-17},
\end{equation}
over the range shown. For integration times $0.1 < \tau < 100$ s, the stability is white frequency noise limited, with a flicker frequency floor below $1\times10^{-16}$. At 1-s integration time $\sigma_y =5.8\times10^{-16}$, a robust result and repeatable over many hours to days of measurement and almost independent of incident power over the range 0.1 mW $\leq P_{inc}\leq$ 0.8 mW. This result is compared with the fractional frequency stability of a single cryogenic oscillator (curve 2) measured from two nominally identical oscillators where the sapphire resonators are exactly the same as those used in this work, but cooled instead with liquid helium in large dewars \cite{Hartnett2006}. Also a comparison is made (curve 3) with the NIST photonic generator where a 10 GHz signal is derived from an optical comb phase-locked to an ultra-stable cavity \cite{Fortier}. 

\begin{figure}[!t]
\centering
\includegraphics[width=3in]{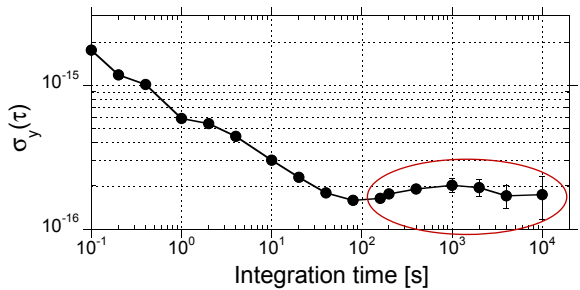}
\caption{(color online) The fractional frequency stability ($\sigma_y(\tau)$) for a single 11.2 GHz crycooled  sapphire oscillator as a function of integration time, from a 12 hour data sample where circled.}
\label{Fig5}
\end{figure}

During the period where these measurements were being made we had a problem with the  liquid/gas helium space in one of the cryostats. It is suspected that air was leaking across an O-ring and contaminating the ultra-pure helium gas. As a result we saw a slow rise of the pressure in the chamber (which is normally kept between 50 kPa and 70 kPa, i.e. below 1 atmosphere) and hence a rise in the temperature of the liquid helium. This has two effects: a) a slow change in the temperature of the cold finger, and  b) the increased pressure reduces the isolation  of the resonator from the environment. As a result,  we saw an increased sensitivity to air-conditioning cycles and the diurnal temperature changes at the resonator resulting in a degradation of the oscillator stability near $\tau = 10^3$ s \cite{Hartnett2010a}. Fig. 5 shows $\sigma_y(\tau)$ up to $\tau = 10^4$ s from a 12 hour sample of relatively quiet beat data where a small hump is apparent around $10^3$ s (circled). This indicates a stability as low as $1\times 10^{-16}$ is achievable once the problem is resolved. 

In conclusion, two nominally identical cryocooled sapphire oscillators where the resonator is cooled to near 6 K with an ultra-low vibration pulse-tube cryocooler and custom built cryostat have been realized and compared. The fractional frequency stability for a single oscillator is $5.8\times 10^{-16}$ for an integration time of 1 s reducing according to a white frequency power law ($\tau^{-1/2}$) to a flicker frequency floor below $1\times 10^{-16}$.

This work was supported by Australian Research Council. The authors thank A. Luiten, E. Ivanov, M. Nagel and E. Kovalchuk for their helpful  suggestions.

\end{document}